\documentclass[a4paper]{article}

\usepackage{INTERSPEECH2021}
\usepackage[utf8]{inputenc}

\title{Zero-Shot Long-Form Voice Cloning with Dynamic Convolution Attention}
\name{Artem Gorodetskii, Ivan Ozhiganov}

\address{
  Azoft Inc., Austin, TX, USA}
\email{artem.gorodetsky@gmail.com, giga@azoft.com}

\begin{document}

\maketitle

\setlength{\footskip}{50pt}
\thispagestyle{plain}
\pagestyle{plain}

\begin{abstract}
    With recent advancements in voice cloning, the performance of speech synthesis for a target speaker has been rendered similar to the human level. However, autoregressive voice cloning systems still suffer from text alignment failures, resulting in an inability to synthesize long sentences. In this work, we propose a variant of attention-based text-to-speech system that can reproduce a target voice from a few seconds of reference speech and generalize to very long utterances as well. The proposed system is based on three independently trained components: a speaker encoder, synthesizer and universal vocoder. Generalization to long utterances is realized using an energy-based attention mechanism known as Dynamic Convolution Attention, in combination with a set of modifications proposed for the synthesizer based on Tacotron 2. Moreover, effective zero-shot speaker adaptation is achieved by conditioning both the synthesizer and vocoder on a speaker encoder that has been pretrained on a large corpus of diverse data. We compare several implementations of voice cloning systems in terms of speech naturalness, speaker similarity, alignment consistency and ability to synthesize long utterances, and conclude that the proposed model can produce intelligible synthetic speech for extremely long utterances, while preserving a high extent of naturalness and similarity for short texts.
\end{abstract}
\noindent\textbf{Index Terms}: Attention mechanism, speech synthesis, text-to-speech, voice cloning, zero-shot learning

\section{Introduction}

Speech synthesis has received significant attention in the research community and has emerged as an essential part of various applications such as dialog systems and voice assistants. The synthesis of natural speech requires training on a considerable amount of transcribed audio data, which complicates the process of model development for a particular speaker \cite{wang2017tacotron, shen2018natural}. In this context, voice cloning systems are most attractive because they can reproduce the voice of a speaker by using only a few samples of recorded speech \cite{jia2018transfer, arik2018neural, paul2020speaker}. This technology has numerous potential applications, for instance, it can be useful for customizing the voice of a digital assistant, translating speech while preserving the speaker’s identity, or even restoring communication ability.

Most voice cloning models consist of three independently trained modules: a speaker encoder, synthesizer and vocoder \cite{jia2018transfer, paul2020speaker, skerry2018towards, cooper2020zero}. The first module represents a speaker-discriminative network that encodes the unique speaker characteristics to a feature vector \cite{wan2018generalized}. The second module is a text-to-speech (TTS) model based on a sequence-to-sequence (seq2seq) architecture \cite{wang2017tacotron, shen2018natural} that learns to convert text sequences into acoustic sequences conditioned on the speaker encoder output. The last module is used to reconstruct time-domain waveforms from the sequence generated by the synthesizer \cite{shen2018natural, prenger2019waveglow, kalchbrenner2018efficient}. Combining all independently trained modules together in a transfer learning configuration allows the system to generalize to previously unseen speakers.

Despite significant advancements in this field, autoregressive voice cloning systems suffer from an inability to synthesize long utterances in a single pass. This inability, which manifests as repeated phonemes, words or even in incomplete synthesis, can be attributed to the limitations of the attention mechanism \cite{battenberg2020location} used to perform the time alignment between the input and output sequences of the TTS model. The original implementation uses a synthesizer based on the Tacotron 2 architecture with a hybrid location-sensitive attention (LSA) \cite{shen2018natural, jia2018transfer} which can accumulate and process attention weights from previous time steps \cite{bahdanau2014neural, chorowski2015attention}. This feature facilitates the synthesis of utterances longer than those used during training. However, the system still suffers from occasional alignment failures and the inability to generalize to extremely long utterances, which severely limits the usage of voice cloning technology.

In recent years, significant progress has been made in addressing the problem of alignment in the context of single-speaker attention-based TTS models \cite{battenberg2020location, zhang2018forward, he2019robust}. Several types of attention mechanisms that directly satisfy the alignment monotonicity have been developed. For example, by using a forward attention mechanism with a transition agent, faster convergence speed and better stability of speech generation can be ensured. However, such system still suffers from frequent skipping and repeating phonemes \cite{zhang2018forward}. To address these problems, stepwise monotonic attention with stronger criteria of alignment monotonicity and completeness has been developed \cite{he2019robust}. However, this approach has drawbacks such as reliance on recursion, complex training procedures and the tendency of punctuation to violate the proposed completeness constraints.

Another group of researchers has attempted to enhance the robustness of LSA by changing the training process of the TTS model \cite{zhu2019pre}. The authors have introduced predefined phoneme durations into the loss function to bias attention learning to the desired direction. The proposed technique exhibits enhanced model stability and generalization performance for longer utterances. The downside of this method is the need to use an external aligner to calculate a priori attention weights.

Other authors have used a modified version of the Gaussian mixture model (GMM) attention introduced in \cite{graves2013generating}. It has been demonstrated that replacement of the exponential function by softplus activation leads to model stabilization \cite{skerry2018towards, kastner2019representation, battenberg2019effective}. A quantitative comparison of various attention mechanisms including LSA and GMM attention is performed in \cite{battenberg2020location}. In addition, the authors have presented two location-relative mechanisms: a modified GMM-based mechanism and a new additive energy-based mechanism, known as Dynamic Convolution Attention (DCA). Both mechanisms outperform the existing concepts of soft attention in terms of robustness and can promote generalization to potentially infinitive-length utterances. In contrast to GMM-based attention, DCA exhibits several advantages, such as strong monotonicity and attention normalization, which are key to ensure robustness and stabilization of alignment. These properties make DCA the preferred choice for zero-shot multispeaker TTS based on the autoregressive architecture.

Other characteristics that limit the use of voice cloning technology are insufficient naturalness and similarity of synthetic speech for unseen speakers. To a certain extent, both characteristics can be significantly improved by increasing the number of unique speakers in the speaker encoder training data. A network, trained on a larger amount of diverse data, tends to produce a more detailed vector representation of the speaker, thereby enhancing the generalization performance of the synthesizer model \cite{jia2018transfer}. In addition, training both synthesizer and vocoder models in the transfer learning configuration can dramatically improve the characteristics of cloned speech. It has been demonstrated that the application of a speaker conditional Wave Recurrent Neural Network (WaveRNN) can help enhance the performance in terms of naturalness and similarity in comparison with that achieved using the conventional WaveRNN vocoder \cite{paul2020speaker}. Notably, the speaker conditional WaveRNN (SC-WaveRNN) provides a high degree of generalization not only for unseen speakers, but also for unseen recording quality, thereby expanding the range of possible applications of the technology.

This study is aimed to develop an autoregressive system capable of reproducing the speech of a target speaker for extremely long utterances in a zero-shot manner, preserving a high extent of naturalness and similarity. Inspired by the performance of the single-speaker Tacotron 1 system with DCA, we extend the application of this location-relative mechanism on the multispeaker Tacotron 2 model. Subsequently, we introduce a set of optimizations and modifications to enhance the consistency of the attention mechanism and increase the similarity to the target speaker. Finally, to enhance the generalization performance, the proposed system makes use of the universal SC-WaveRNN vocoder and the speaker encoder pretrained on a significantly larger set of speakers than that previously reported \cite{jia2018transfer, paul2020speaker, cooper2020zero}. Our results based on objective and subjective evaluations demonstrate that compared to the baseline system \cite{jia2018transfer}, the proposed model can produce speech with higher speaker similarity with the ability to generalize to extremely long utterances.

Recently, several non-autoregressive flow-based architectures for multispeaker TTS have been proposed \cite{casanova2021sc, casanova2021yourtts}. These models can perform zero-shot voice cloning and potentially generalize to long utterances. However, these works mostly focus on inference speed or multilingual approaches rather than producing natural speech using long texts. In contrast, our system utilizes soft attention and, to our knowledge, represents the first autoregressive voice cloning framework capable of synthesizing natural speech using extralong utterances.

The rest of the paper is organized as follows. Section 2 describes the components of the multispeaker TTS system, including a formulation of an additive energy-based attention mechanism. In Section 3 we introduce the proposed model for long-form voice cloning with zero-shot speaker adaptation. Section 4 describes the experimental setup and training procedures. Section 5 introduces the results and discussion. Finally, conclusions are reported in Section 6.

\section{Multispeaker TTS Architecture}
\subsection{Neural speaker encoder}

The speaker encoder is used to obtain a fixed-dimensional embedding, known as the d-vector, from audio samples of the speaker \cite{heigold2016end, variani2014deep}. Without retraining the TTS system, the embedding vector is fed to the synthesizer and vocoder to reproduce the voice of the target speaker. Since the TTS system is fully conditioned on the speaker encoder, its ability to generalize to new speakers is crucial for all parts of the framework. To ensure generalization, the network is trained on a speaker verification task using a generalized end-to-end (GE2E) loss function and audio records of thousands of speakers \cite{wan2018generalized}. The application of the GE2E loss allows the construction of an embedding space in which d-vectors from the same speaker exhibit a high cosine similarity, while d-vectors from different speakers are located far from one another and have a substantially lower similarity value \cite{jia2018transfer, paul2020speaker}. 

The model represents a recurrent neural network (RNN), which encodes a sequence of mel-spectrogram frames extracted from speech records into embedding vectors. During training, the batch is composed of $N$ $\times$ $M$ mel-spectrograms from $N$ speakers with $M$ utterances per speaker, so that each feature vector $\boldsymbol{z}_{nm}$ ($1\leq n \leq N$ and $1 \leq m \leq M$) from the batch represents features computed from utterance $m$ of speaker $n$. After feeding the vector $\boldsymbol{z}_{nm}$ for each utterance to the speaker encoder, the final set of embeddings $\boldsymbol{e}_{nm}$ is calculated as the L2-normalization of the model output. The obtained d-vectors $\boldsymbol{e}_{nm}$ and speaker centroids $\boldsymbol{u}_{k}$ ($1 \leq k \leq N$) define the similarity matrix $\boldsymbol{S}_{nm,k}$ through scaled cosine similarity with learnable parameters $\omega$ and $b$ \cite{wan2018generalized}:
\vspace{-0.5em}
\begin{equation}
    \boldsymbol{S}_{nm,k}=\begin{cases}
    \omega\cdot\cos(\boldsymbol{e}_{nm}, \boldsymbol{u}_{n}^{-m})+b, & \text{if } k = n\\
    \omega\cdot\cos(\boldsymbol{e}_{nm}, \boldsymbol{u}_{k})+b, & \text{otherwise}
    \end{cases},
    \label{eq1}
    \vspace{-0.5em}
\end{equation}
where
\vspace{-0.5em}
\begin{equation}
    \boldsymbol{u}_{n}^{-m} = \frac{1}{M-1}\sum_{i=1, i \neq m}^{M} \boldsymbol{e}_{ni}, \quad \boldsymbol{u}_{k}=\frac{1}{M}\sum_{i=1}^{M} \boldsymbol{e}_{ki}.
    \label{eq2}
    \vspace{-0.5em}
\end{equation}
The probability that the embedding vector $\boldsymbol{e}_{nm}$ belongs to a particular speaker is determined by applying the softmax function to $\boldsymbol{S}_{nm,k}$ for $k=1, ..., N$. Subsequently, the GE2E loss $L_{GE2E}$ is defined as the cross-entropy summed over the similarity matrix for each d-vector:
\vspace{-0.5em}
\begin{equation}
    L_{GE2E} = \sum_{n,m}^{N,M}[-\boldsymbol{S}_{nm,n} + \log(\sum_{k}^{N}\exp(\boldsymbol{S}_{nm,k}))].
    \label{eq3}
    \vspace{-0.5em}
\end{equation}

\subsection{Multispeaker Tacotron 2}

\begin{figure}[t]
  \centering
  \includegraphics{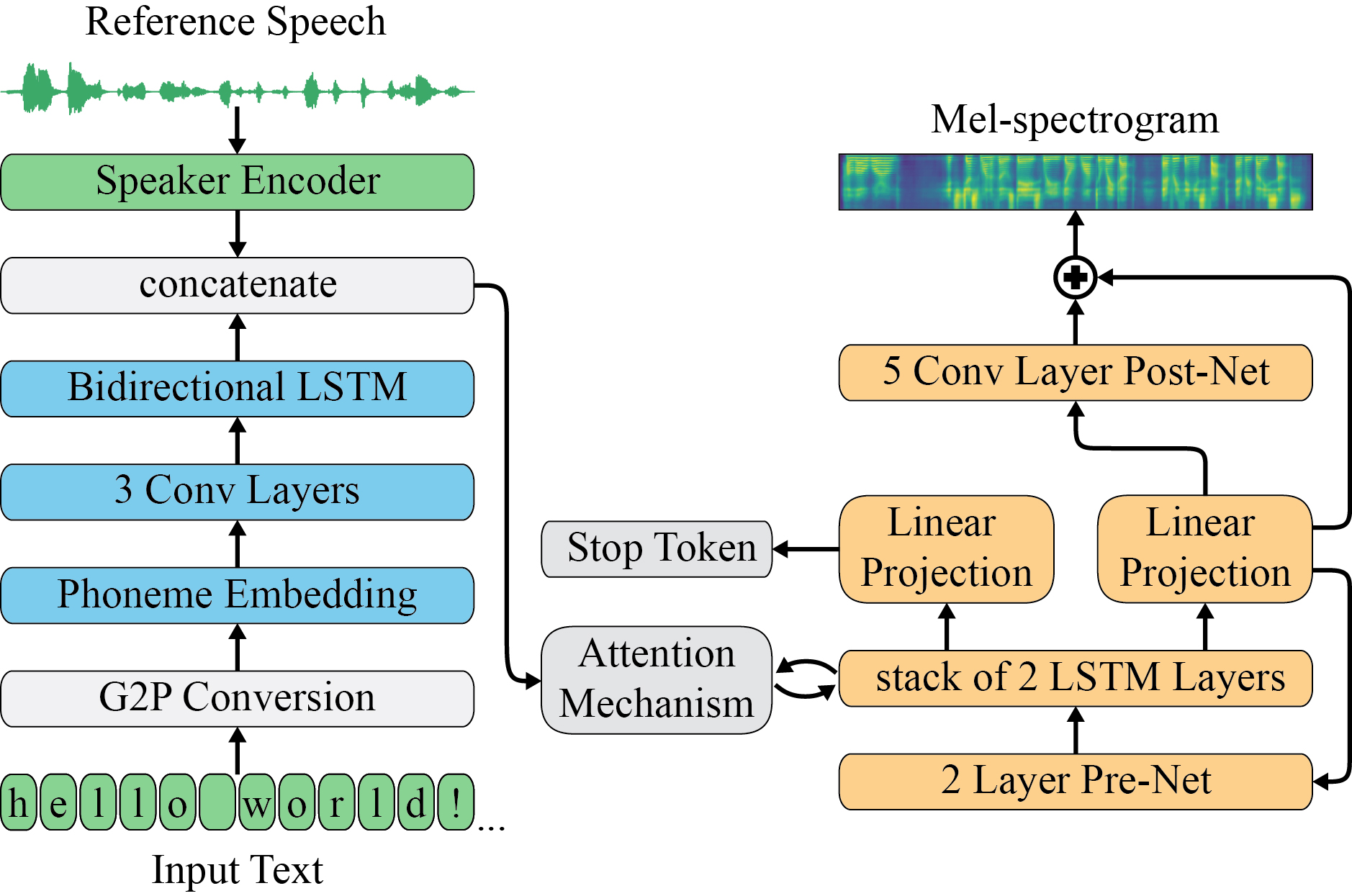}
  \caption{Block diagram of the multispeaker Tacotron 2 architecture. The model takes character sequence and speaker embedding as inputs and produces the mel-spectrogram. The model architecture is shown without the vocoder module. G2P conversion represents a grapheme-to-phoneme conversion pipeline, which transforms a sequence of characters to a sequence of phonemes.}
  \label{fig1} 
  \vspace{-2.0em}
\end{figure}

Tacotron 2 is a seq2seq network composed of encoder and decoder modules with an attention mechanism \cite{shen2018natural}. Figure \ref{fig1} shows the block diagram of a multispeaker version of the Tacotron 2 model \cite{jia2018transfer}. The encoder \eqref{eq4} converts the input text mapped to a phoneme sequence of length $L$, $\{x_{j}\}_{j=1}^{L}$, into a hidden representation $\{\boldsymbol{h}_{j}\}_{j=1}^{L}$. To generate these features, phonemes are represented by phoneme embeddings, which are passed through a stack of 3 convolution layers and a single bidirectional \cite{schuster1997bidirectional} LSTM \cite{hochreiter1997long} layer. Then the obtained sequence $\{\boldsymbol{h}_{j}\}_{j=1}^{L}$ is extended by concatenation with the d-vector $\boldsymbol{e}_{nm}$, forming a new hidden representation for a particular speaker, $\{\boldsymbol{h}_{j}^{nm}\}_{j=1}^{L}$:
\vspace{-0.5em}
\begin{equation}
    \{\boldsymbol{h}_{j}\}_{j=1}^{L} = Encoder(\{x_{j}\}_{j=1}^{L}),
    \label{eq4}
\end{equation}
\begin{equation}
    \{\boldsymbol{h}_{j}^{nm}\}_{j=1}^{L} = \{[\boldsymbol{h}_{j}; \boldsymbol{e}_{nm}]\}_{j=1}^{L}.
    \label{eq5}
\end{equation}
The decoder iteratively transforms the encoder output into a sequence of mel-spectrogram frames by using an attention mechanism. The attention network \eqref{eq6} functions as a bridge between the encoder and decoder modules to produce $\boldsymbol\alpha_{i}$, i.e., the encoder-decoder time alignments at decoder step $i$. Considering the time alignments $\boldsymbol\alpha_{i}$ as weights, the attention context vector, $\boldsymbol{c}_{i}$, is computed as the weighted average of the states $\{\boldsymbol{h}_{j}^{nm}\}_{j=1}^{L}$. The further decoding process is organized as follows. First, the prediction from the previous decoder timestep, $\boldsymbol{y}_{i-1}$, is passed through a 2-layer pre-net. The decoder RNN represented by a stack of 2 unidirectional LSTM layers uses the concatenation of the pre-net output $\boldsymbol{y}_{i-1}^{*}$ and the context vector $\boldsymbol{c}_{i-1}$ to compute the current hidden state $\boldsymbol{d}_{i}$. Next, a fully connected layer processes the concatenation of the decoder RNN output, $\boldsymbol{d}_{i}$, and the attention context, $\boldsymbol{c}_{i}$, to predict the next spectrogram frame, $\boldsymbol{y}_{i}$:
\vspace{-0.5em}
\begin{equation}
    \boldsymbol\alpha_{i} = Attention(\boldsymbol{d}_{i}, ... ), \quad \boldsymbol{c}_{i} = \sum_{j=1}\alpha_{ij}\boldsymbol{h}_{j}^{nm},
    \label{eq6}
    \vspace{-0.5em}
\end{equation}
\begin{equation}
    \boldsymbol{y}_{i-1}^{*} = PreNet(\boldsymbol{y}_{i-1}),
    \label{eq7}
\end{equation}
\begin{equation}
    \boldsymbol{d}_{i} = DecoderRNN(\boldsymbol{d}_{i-1}, [\boldsymbol{y}_{i-1}^{*}; \boldsymbol{c}_{i-1}]),
    \label{eq8}
\end{equation}
\begin{equation}
    \boldsymbol{y}_{i} = f([\boldsymbol{d}_{i}; \boldsymbol{c}_{i}]).
    \label{eq9}
\end{equation}
Other arguments in \eqref{eq6} are determined by the nature of the attention mechanism. Moreover, the concatenation of the decoder RNN hidden state and the context vector is used to predict the probability of a “stop token”, allowing one to determine when the frame generation process must be terminated. Finally, to enhance the reconstruction quality, the predicted mel-spectrogram is fed to a convolutional post-net, the output of which is added to the prediction via a residual connection. During training, the model is optimized to minimize a combination of the L1 and L2 losses on the mel-spectrograms obtained before and after the post-net \cite{shen2018natural, jia2018transfer}. Binary cross-entropy is used as a “stop token” loss.

\subsection{Energy-based attention mechanisms}

The original Tacotron 2 system utilizes the hybrid LSA, which belongs to a family of energy-based mechanisms. These mechanisms use a multilayer perceptron to compute attention energies, $\boldsymbol{E}_{i}$, that are transformed into attention weights, $\boldsymbol\alpha_{i}$, by the softmax function. A formulation of this approach adapted for the multispeaker Tacotron 2 is expressed as \eqref{eq10} \cite{battenberg2020location}:
\begin{equation}
    \begin{aligned}
        {E}_{i,j} = \boldsymbol\upsilon ^{T} \tanh(& W\boldsymbol{d}_{i} +V\boldsymbol{h}_{j}^{nm}+ U\boldsymbol{f}_{i,j} \\
        & + T\boldsymbol{g}_{i,j}+b)+\boldsymbol{p}_{i,j},
       \label{eq10}
    \end{aligned}
\end{equation}
\begin{equation}
    \boldsymbol\alpha_{i} = Softmax(\boldsymbol{E}_{i}),
    \label{eq11}
\end{equation}
\begin{equation}
    \boldsymbol{f}_{i} = \mathcal{F} \ast \boldsymbol\alpha_{i-1},
    \label{eq12}
\end{equation}
\begin{equation}
    \boldsymbol{g}_{i} = \mathcal{G}(\boldsymbol{d}_{i}) \ast \boldsymbol\alpha_{i-1}, \quad \mathcal{G}(\boldsymbol{d}_{i})=V_{\mathcal{G}}\tanh( W_{\mathcal{G}}\boldsymbol{d}_{i} + b_{\mathcal{G}}),
    \label{eq13}
\end{equation}
\begin{equation}
    \boldsymbol{p}_{i} = \log(\mathcal{P} \ast \boldsymbol\alpha_{i-1}).
    \label{eq14}
\end{equation}
The hybrid LSA attention contains content-based terms, $W\boldsymbol{d}_{i}$ and $V\boldsymbol{h}_{j}^{nm}$, that perform query and key comparison, as well as the location-sensitive term, $U\boldsymbol{f}_{i,j}$, that takes into account the time alignments produced in the previous timestep by using a set of static convolution filters, as shown in \eqref{eq12} \cite{chorowski2015attention}. Two other terms, $T\boldsymbol{g}_{i,j}$ and $\boldsymbol{p}_{i,j}$, in combination with the location-sensitive term, $U\boldsymbol{f}_{i,j}$, are used in DCA \cite{battenberg2020location}. The $T\boldsymbol{g}_{i,j}$ term applies a set of learned dynamic filters computed from the last decoder RNN hidden state to the previous alignments, as indicated in \eqref{eq13}. The $\boldsymbol{p}_{i,j}$ term uses a single fixed prior filter based on a beta-binomial distribution to stimulate forward movement of the attention. In contrast to the Tacotron 1 architecture \cite{wang2017tacotron, battenberg2020location}, our implementation of Tacotron 2 \cite{shen2018natural} does not contain an attention RNN. Consequently, the content-based term $T\boldsymbol{g}_{i,j}$ and the dynamic convolution filters are applied directly to the decoder RNN output, as indicated in \eqref{eq10} and \eqref{eq13}, respectively.

\subsection{Speaker conditional WaveRNN vocoder}

The family of WaveRNN networks provides a simple and powerful tool to realize the sequential modeling of high-fidelity audio \cite{kalchbrenner2018efficient, valin2019lpcnet}. Our implementation is based on an alternative version proposed in \cite{paul2020speaker}, known as the SC-WaveRNN vocoder. The network converts the synthesized mel-spectrogram into time-domain waveform samples by using the speaker embedding vector as additional information. This feature renders the vocoder universal, allowing the system to control the characteristics of synthesized speech even for unseen speakers and recording conditions.

In contrast to the original SC-WaveRNN, which produces a mixture of logistic distributions at each time step \cite{oord2018parallel}, the proposed model outputs a categorical distribution with a softmax layer. We apply a $\mu$-law encoding \cite{itu1972recommendation} to the input data and quantize it to 512 possible values. Subsequently, the inverse $\mu$-law transformation is applied to reconstruct the predicted signal.
During training, the network is optimized to maximize the log-likelihood of the data.

\section{Zero-Shot Long-Form Voice Cloning}

The speaker encoder and Dynamic Convolution Attention are two key components used to develop an autoregressive TTS system capable of synthesizing voices of new speakers and using long sentences or even paragraphs. The system does not require retraining of the model or its parts and relies on a d-vector computed by the speaker encoder from a few seconds of reference speech. This model adaptation technique is commonly known as zero-shot learning.
\begin{figure*}[t]
  \centering
  \includegraphics{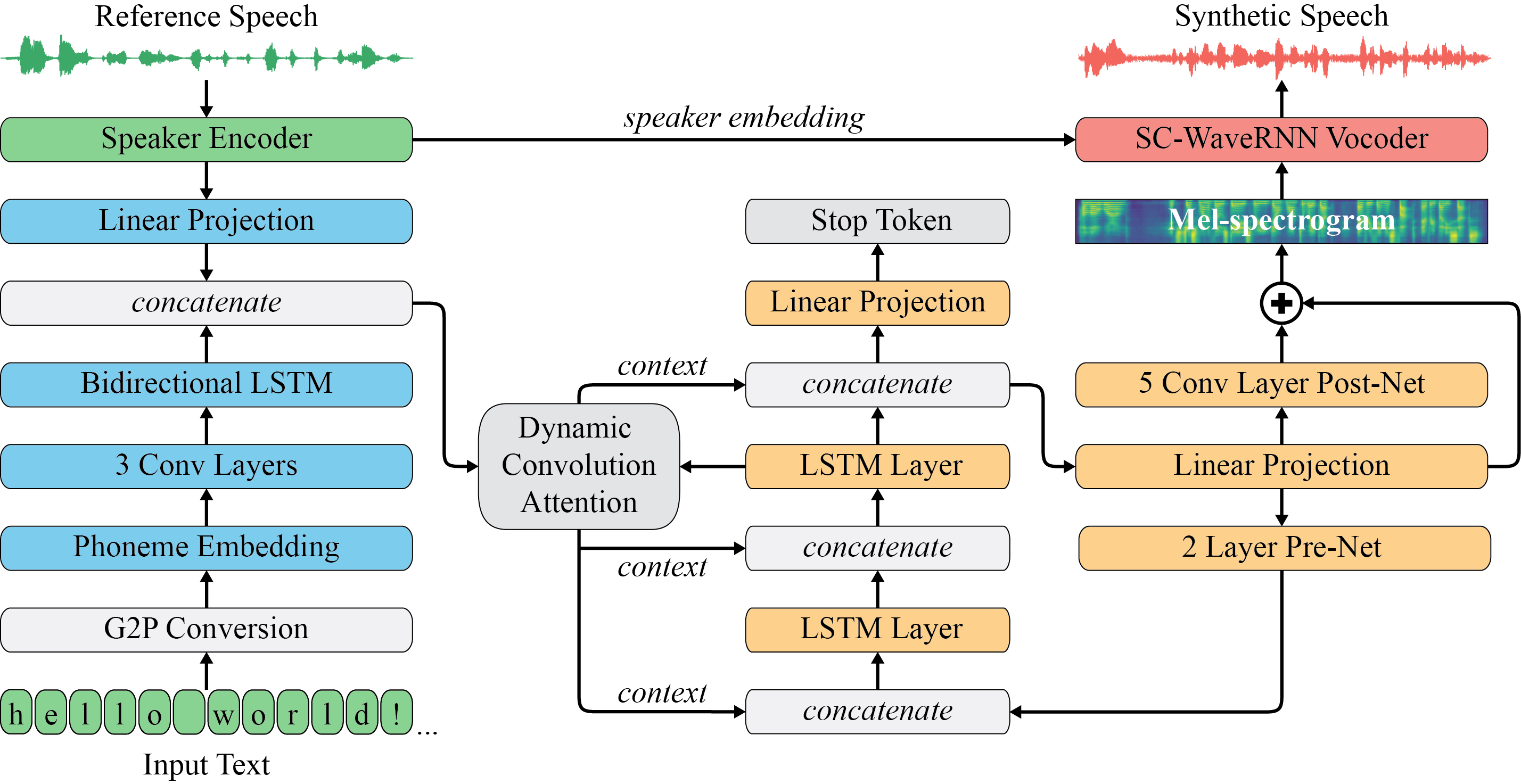}
  \caption{Block diagram of the proposed multispeaker system for zero-shot long-form TTS. The multispeaker Tacotron 2 model takes the character sequence and speaker embedding vector as inputs and outputs the mel-spectrogram. The SC-WaveRNN vocoder conditioned on the speaker encoder output is used to transform the mel-spectrogram into audio waveforms. Context and G2P denote the attention context vector and grapheme-to-phoneme pipeline, respectively.}
  \label{fig2}
  \vspace{-2.0em}
\end{figure*}

The proposed system is based on three independently trained models, as described in Section 2: the neural speaker encoder, multispeaker Tacotron 2 and universal SC-WaveRNN vocoder. The d-vector computed by the speaker encoder in an utterance-wise manner is used for conditioning the mel-spectrogram synthesized by Tacotron 2 and time-domain waveforms generated by the vocoder. The block diagram of the proposed system is shown in Figure \ref{fig2}. We use the DCA instead of the LSA mechanism, allowing the model to generalize to extralong utterances. Moreover, we introduce other architectural changes in the multispeaker Tacotron 2, thereby enhancing the quality of the alignment process: (a) the speaker embedding vector is passed through an additional linear layer to stimulate the extraction of more meaningful speaker characteristics; (b) a skip connection represented by the concatenation of the first decoder LSTM output with the attention context vector is added, as shown in Figure 2; (c) the previous time step context vector, $\boldsymbol{c}_{i-1}$, is used to predict the next mel-spectrogram frame in \eqref{eq9}. In addition to the regularizations proposed for the original single-speaker Tacotron 2 \cite{shen2018natural}, we apply dropout \cite{srivastava2014dropout} with probability 0.1 to the input of the dynamic convolution filters \eqref{eq13} and increase the zoneout \cite{krueger2016zoneout} probability for the second decoder LSTM layer to 0.15. In practice, it was found that all of these changes result in improved alignment consistency.

\section{Experimental Setup}

We experimentally compare several implementations of the multispeaker TTS system. We consider the model proposed in \cite{jia2018transfer} as the default baseline system (Tacotron2-LSA-40) and replace WaveNet \cite{oord2016wavenet} with the SC-WaveRNN vocoder. In other implementations, we change the speaker encoder network, attention mechanism and architecture of the Tacotron 2 model. In this manner, the following TTS systems and their components are developed:

\begin{itemize}
    \item \textit{Tacotron2-LSA-40}: 40-channel speaker encoder, multispeaker Tacotron 2 with LSA, SC-WaveRNN;
    \item \textit{Tacotron2-LSA-80}: 80-channel speaker encoder, multispeaker Tacotron 2 with LSA, SC-WaveRNN;
    \item \textit{Tacotron2-DCA-80}: 80-channel speaker encoder, multispeaker Tacotron 2 with DCA, SC-WaveRNN;
    \item \textit{Proposed model}: 80-channel speaker encoder, modified multispeaker Tacotron 2 with DCA, as described in Section 3, SC-WaveRNN.
\end{itemize}

The speaker encoder is trained in two configurations that take as inputs 40-channel and 80-channel mel-spectrograms extracted from audio frames with a width and step of 25 ms and 10 ms, respectively. Both models consist of 3 LSTM layers with 768 cells followed by a linear projection to 256 dimensions without an activation function. Training is conducted on a data corpus based on four public sets: LibriSpeech \cite{panayotov2015librispeech} (train-other-500 subset), VoxCeleb 1 \cite{nagrani2017voxceleb} (development subset), VoxCeleb 2 \cite{chung2018voxceleb2} (development subset) and Common Voice 5.1 \cite{ardila2019common} (only English data), containing a total of 1.85 M utterances from 38,695 speakers. The union of two LibriSpeech development subsets (dev-clean and dev-other subsets) is used for validation. The audio samples are processed by a voice activity detector (VAD) to remove prolonged speech silence sections. The final training dataset consists of partial utterances with a fixed duration of 1.6 s sampled from the processed complete utterances. During training, each batch contains $N = 64$ speakers and $M = 10$ utterances per speaker. We use the Adam optimizer with $\beta_{1}$ = 0.9, $\beta_{2}$ = 0.999, $\varepsilon$ = $10^{-8}$ for 1.5 M steps with a gradient clipping threshold of 3 and an initial learning rate of $10^{-4}$ that is decreased by half after 1 M steps. During inference, each utterance is divided into 1.6 s windows with 50$\%$ overlap, and the final utterance embedding is computed as the L2-normalized average over the d-vectors of each individual window. 

The Tacotron 2 models are trained using two subsets of the LibriTTS corpus \cite{zen2019libritts} (train-clean-100 and train-clean-300 subsets), containing approximately 245 h of speech from 1,151 speakers. The development subset of LibriTTS with clean speech (dev-clean subset) is used for validation. The models input consists of phoneme sequences produced by a grapheme-to-phoneme conversion pipeline and speaker embeddings extracted from speech samples. The target spectrogram features represent 80-channel mel-spectrogram frames computed from a 16 kHz audio using a frame size of 50 ms with a 12.5 ms step. All models are implemented with a reduction factor, $r$, of 2, meaning that for each decoder step, two spectrogram frames are predicted. For LSA, we utilize 32 static filters of length 31, while for DCA, we apply 8 static and 8 dynamic filters and one prior filter with the same lengths and parameters as described in \cite{battenberg2020location}. For both attention mechanisms, a 128-dimensional hidden representation is used.

Each Tacotron 2 model is trained with a batch size of 64 for 300k iterations, using the Adam optimizer with $\beta_{1}$ = 0.9, $\beta_{2}$ = 0.999, $\varepsilon$ = $10^{-6}$ and weight decay of $10^{-6}$. The models for the Tacotron2-LSA-40, Tacotron2-LSA-80 and Tacotron2-DCA-80 systems are trained with a gradient clipping threshold of 1 and a learning rate of $10^{-3}$ that exponentially decays to $10^{-5}$ after 50k steps \cite{shen2018natural}. To train the proposed model, we apply a gradient clipping threshold of 0.05 and an initial learning rate of $10^{-3}$ that is reduced by 50$\%$ after 10k, 20k, 40k, 60k, 100k, 150k, 200k and 250k iterations. We found that these settings also lead to improvement of alignment.

To convert the mel-spectrograms into audio samples, we separately train SC-WaveRNN for each multispeaker Tacotron 2 model. All vocoders are trained for 1020 epochs on ground-truth-aligned predictions of a synthesizer network using speaker embedding vectors computed in an utterance-wise manner. We use the Adam optimizer with $\beta_{1}$ = 0.9, $\beta_{2}$ = 0.999, $\varepsilon$ = $10^{-8}$, a batch size of 32, a gradient clipping threshold of 4 and a learning rate of $10^{-4}$ that is reduced by half after 340, 510, 680 and 850 epochs.

All neural networks are implemented using the PyTorch framework and trained on a single NVIDIA RTX 3090 GPU with 24 GB memory.

\section{Results and Discussion}

\subsection{Subjective evaluation}

We evaluate the TTS models by considering mean opinion score (MOS) naturalness and similarity judgments produced by a pool of raters on Amazon Mechanical Turk. We construct an evaluation set containing 5 male and 5 female speakers from the test-clean subset of LibriTTS. For each speaker, 10 phrases and 10 utterances with durations longer than 5 s are randomly selected. Synthetic speech is generated in a zero-shot manner using prepared texts and speaker embeddings computed from the selected utterances. Test scores range from 1 to 5 in half-point increments, where 5 represents “perfectly natural speech” or “perfectly similar voice” in the naturalness and similarity tests, respectively. To evaluate speech similarity, each synthesized utterance is paired with its ground truth utterance, which is used to compute the speaker embedding vector. The evaluation results are shown in Table \ref{tab1}. We see that for these utterances, all models achieve equivalent naturalness and similarity MOS scores around 4.0 and 3.9, respectively.

\begin{table}[th]
    \vspace{-0.5em}
    \caption{Mean opinion scores (MOS) with 95$\%$ confidence intervals. All speakers and texts were unseen during training.}
    \label{tab1}
    \centering
    \setlength{\tabcolsep}{5pt}
    \begin{tabular}{ c c c }
        \toprule
        \textbf{System} & \textbf{Naturalness} & \textbf{Similarity} \\
        \midrule
        Ground truth     & 4.47 $\pm$ 0.04 & 4.41 $\pm$ 0.04 \\ 
        Tacotron2-LSA-40 & 4.02 $\pm$ 0.05  & 3.93 $\pm$ 0.05 \\
        Tacotron2-LSA-80 & 3.99 $\pm$ 0.05 & 3.89 $\pm$ 0.06 \\
        Tacotron2-DCA-80 & 3.99 $\pm$ 0.05 & 3.90 $\pm$ 0.06 \\
        Proposed model   & 4.01 $\pm$ 0.05 & 3.91 $\pm$ 0.06 \\
        \bottomrule
    \end{tabular}
    \vspace{-1em}
\end{table}

\subsection{Objective evaluation}

For objective evaluation, we construct a set based on the test-clean subset of the LibriTTS corpus. The constructed set contains data from 35 speakers (17 male and 18 female voices) with 10 randomly chosen utterances for each speaker (with duration $\geq$ 5 s). We perform zero-shot voice cloning for each utterance using randomly selected texts from the same subset of LibriTTS. To obtain statistically reliable results, each experiment is performed 30 times, so 10,500 utterances are synthesized to evaluate each implemented TTS system. 

To evaluate the speaker similarity between cloned and reference speech, we calculate the speaker verification equal error rates (SV-EER). We enroll the voices of 35 real speakers from the constructed evaluation set by using all 10 utterances per speaker. The 80-channel speaker encoder is used as a speaker verification system to score the similarity between two utterances based on the cosine similarity of their d-vectors. The SV-EERs are determined by paring each synthesized utterance with each enrollment speaker \cite{jia2018transfer}. Moreover, we calculate the average cosine similarities ($S_{C}$) between the embeddings extracted from synthetic speech and their ground truth utterances as an additional measure of the voice cloning quality.

To estimate the degree of alignment consistency, we measure an attention diagonal score computed as 
\vspace{-0.5em}
\begin{equation}
    A_{S} = \frac{1}{L}\sum_{j=1}^{L}{max}_{1 \leq i \leq \frac{T}{r}}(\alpha_{ij}),
    \label{eq17}
    \vspace{-0.5em}
\end{equation}
where $r$ is the reduction factor; $L$ is the length of the phoneme sequence; and $T$ is the number of frames in the predicted mel-spectrogram. In contrast to the focus rate \cite{ren2019fastspeech}, the proposed attention score averages the maximum values of the attention matrix along the encoder steps rather than the decoder timesteps. We found this approach to be more sensitive to alignment quality.

Moreover, we compute the average Mel cepstral distortion (MCD) \cite{kubichek1993mel} coefficients to evaluate the synthetic voices. To obtain this objective metric, the utterances are synthesized using the texts of their ground truth samples. MCD values are calculated as the root square error on 13-dimensional mel-frequency cepstral coefficients obtained from 80-channel mel-spectrograms \cite{weiss2021wave}. To align features extracted from synthetic and real speech, we apply the dynamic time warping algorithm \cite{muller2015fundamentals}.

\begin{table}[th]
    \vspace{-0.5em}
    \caption{Results of objective evaluation. $S_{C}$ and $A_{S}$ values are shown with their standard deviations. All speakers and texts were unseen during training.}
    \label{tab2}
    \centering
    \setlength{\tabcolsep}{2.75pt}
    \begin{tabular}{ c c c c c}
        \toprule
        \textbf{System} & 
        \textbf{SV-EER} & 
        \boldsymbol{$S_{C}$} & 
        \boldmath$A_{S}$ & 
        \textbf{MCD} \\
        \midrule
        Ground truth & 0.23$\%$ & 83 $\pm$ 9$\%$ & – & – \\ 
        Tacotron2-LSA-40 & 1.19$\%$ & 80 $\pm$ 5$\%$ & 0.52 $\pm$ 0.03 & 4.57 \\ 
        Tacotron2-LSA-80 & 0.95$\%$ & 82 $\pm$ 5$\%$ & 0.59 $\pm$ 0.03 & 4.55 \\
        Tacotron2-DCA-80 & 0.92$\%$ & 82 $\pm$ 5$\%$ & 0.51 $\pm$ 0.02 & 4.58 \\
        Proposed model & 0.85$\%$ & 82 $\pm$ 5$\%$ & 0.68 $\pm$ 0.03 & 4.56 \\
        \bottomrule
    \end{tabular}
\end{table}

The results of the objective evaluation are shown in Table \ref{tab2}, which compares the performance of the developed multispeaker TTS systems. All models achieve similar MCD values, consistent with the results of the subjective naturalness test (see Table \ref{tab1}). However, voice cloning systems differ dramatically from one another in terms of SV-EER and alignment consistency. The SV-EER is significantly better for all TTS systems with the 80-channel speaker encoder and reaches the lowest value for the proposed model with a relative improvement of 28.6$\%$. This finding demonstrates that increasing the input dimension in the GE2E speaker verification task leads to a more meaningful speaker representation, thereby enhancing the generalization properties of the entire TTS system. At the same time, the average $S_{C}$ increases by only two percent. Nevertheless, analysis of the $S_{C}$ values indicates that, on average, the synthesized speech tends to be very close to the target speaker with approximately the same $S_{C}$ value as for the ground truth samples.

It is noteworthy that improving the quality of d-vectors leads to a higher degree of alignment consistency in the case of LSA (see Table \ref{tab2}). Because the content-based term directly processes speaker embeddings, as indicated in \eqref{eq10}, their quality can influence the processes of query and key comparisons. On the other hand, a more detailed speaker representation results in improved convergence across different speakers, which likely affects the quality of alignment for both types of the considered attention mechanisms. Nevertheless, compared with the baseline system, simply replacing LSA with DCA does not enhance $A_{S}$ but rather deteriorates the alignment quality. In contrast to Tacotron2-DCA-80, the proposed model exhibits a considerably higher $A_{S}$ (with a relative improvement of 33.3$\%$) and, therefore, has a more consistent attention mechanism. This fact underlines the importance of the changes introduced to the synthesizer architecture and training hyperparameters.

\begin{figure}[t]
  \centering
  \includegraphics{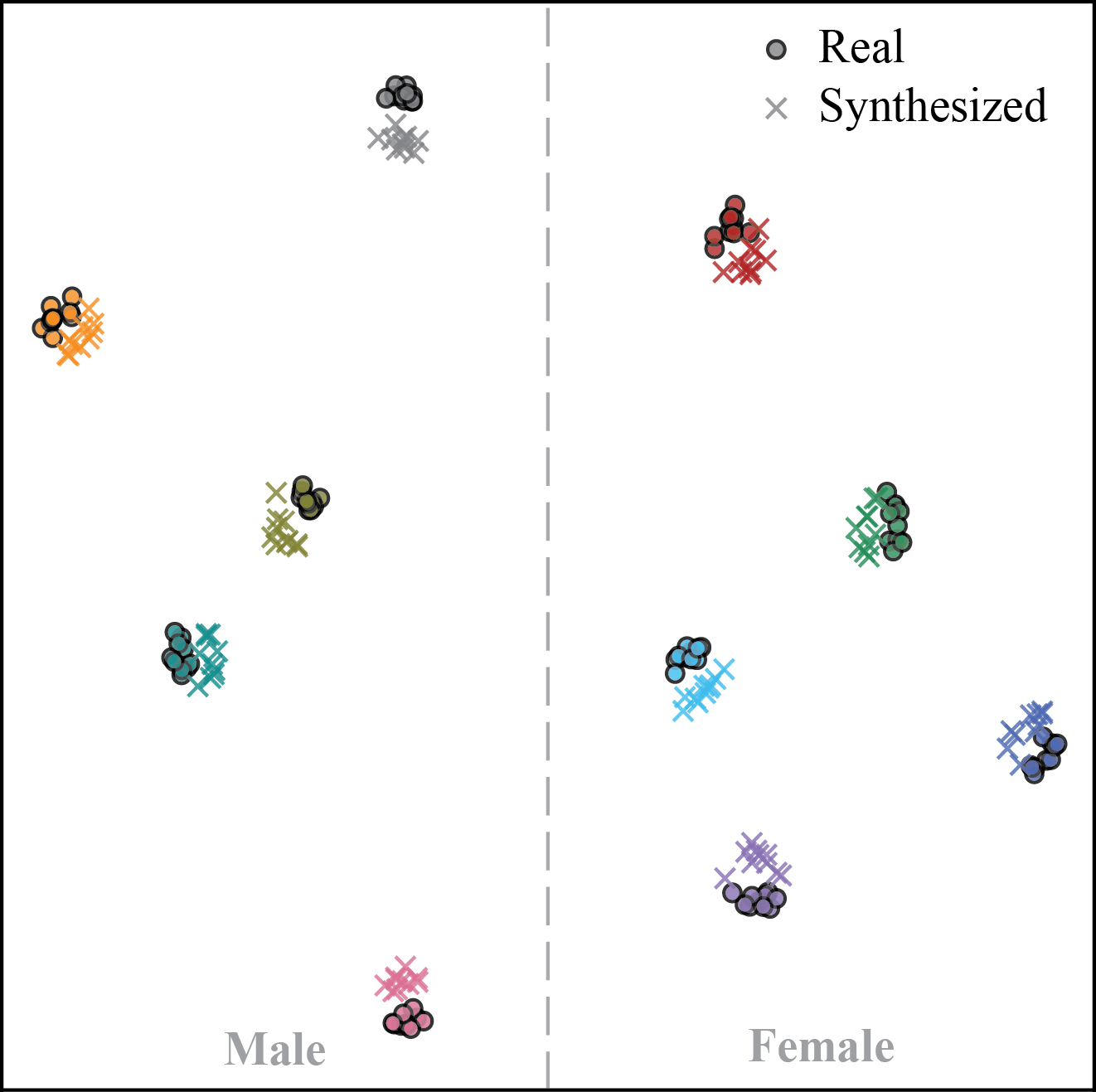}
  \caption{Visualization of speaker embeddings of real and synthesized utterances using t-SNE. Different colors represent different speakers. Male speakers are on the left and female speakers are on the right side of the figure. Voice cloning was performed in the zero-shot manner using the proposed model. All speakers and texts were unseen during training.}
  \label{fig3}
  \vspace{-2.0em}
\end{figure}

In addition, we estimate the difficulty of distinguishing synthetic speech from real speech by using the procedure described in \cite{jia2018transfer}. Specifically, we calculate the equal error rates on a 70-voice discriminative task, in which half of the enrolled speakers are real and the other half correspond to synthetic versions. For all models we obtain values of approximately 1.6$\%$. Therefore, although synthesized speech tends to be close to the target speaker, it can be distinguished from real speech, as demonstrated in Figure \ref{fig3}. The plot shows the visualization of d-vectors extracted from utterances with real and synthesized speech by using t-distributed stochastic neighbor embedding (t-SNE) \cite{van2008visualizing}. We see that the real and synthesized utterances are located very close to one another when they belong to the same speaker, but synthetic speech still tends to form distinct clusters.

\subsection{Generalization to long utterances}

We evaluate the ability of the developed models to synthesize speech using long utterances. The set of utterances is extracted from three chapters of the Harry Potter novels. We use 1036 texts with lengths ranging from 59 to 1653 characters. The Vosk\footnote{https://alphacephei.com/vosk/} model based on the Kaldi-active-grammar repository\footnote{https://github.com/daanzu/kaldi-active-grammar} is used to produce transcripts of the synthesized utterances. To estimate speech intelligibility, we calculate the character error rate (CER) between the produced transcripts and the ground truth transcripts. Each phrase is synthesized in a zero-shot manner for each of the 35 speakers from the abovementioned evaluation set, so 36,260 utterances are generated per evaluation. In addition to the CER, we propose a new objective metric, the silence rate, which reflects the fluency of synthetic speech. The silence rate is calculated as the percentage of silent areas in the utterance determined by the VAD model.
\begin{figure}[t]
  \centering
  \includegraphics{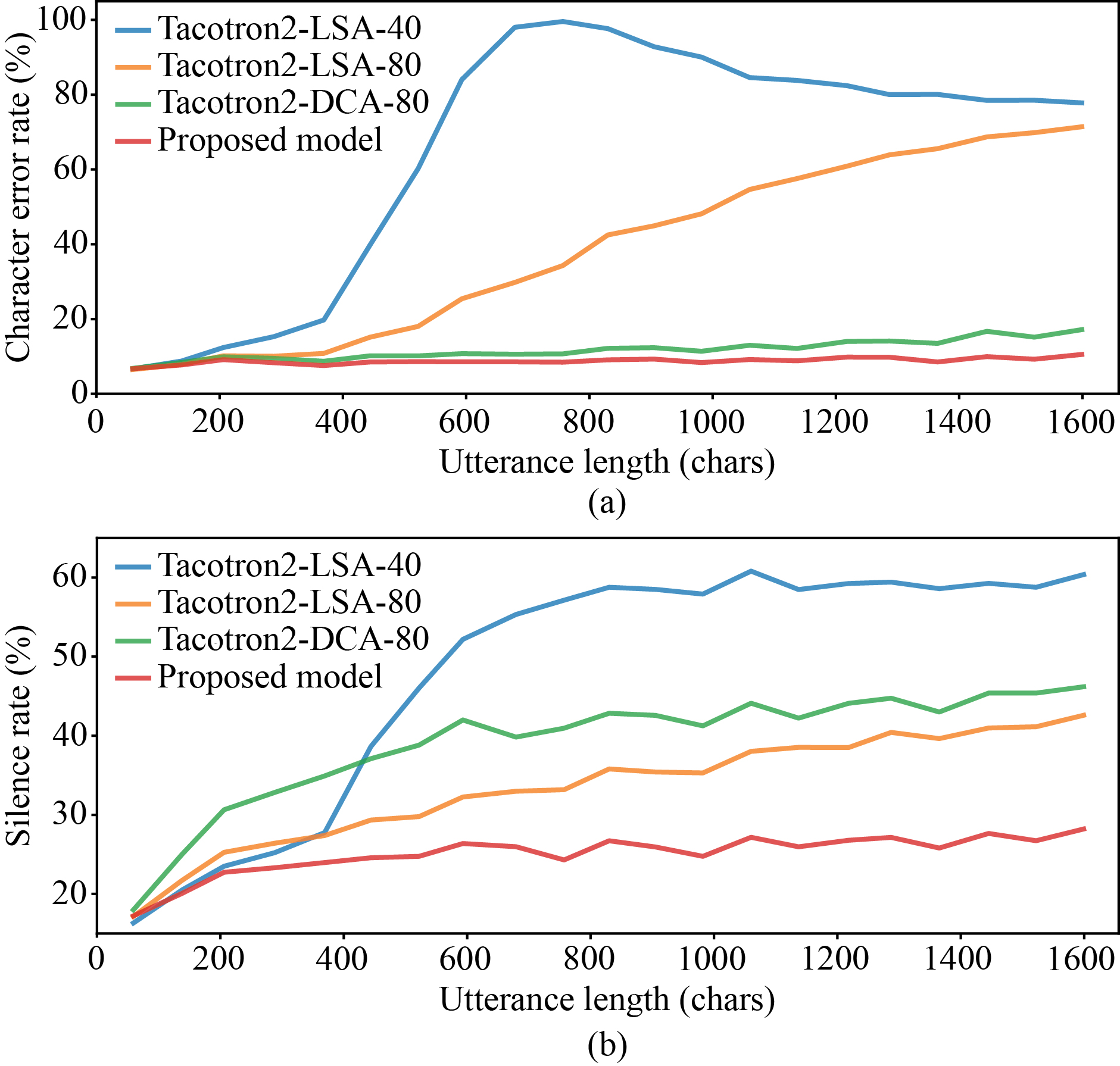}
  \caption{Dependencies of averaged (a) character error rate and (b) silence rate on utterance length. The plots reflect the robustness of the developed models in terms of the utterance length. All speakers and texts were unseen during training.}
  \label{fig4}
  \vspace{-2.0em}
\end{figure}

The averaged CER and silence rate as functions of the utterance length are shown in Figure \ref{fig4}. We see that for the baseline system, the CER increases dramatically when the length of the text exceeds the maximum training length. However, for the model with the 80-channel encoder, a more gradual increase in the error is observed. For both models with LSA, the percentage of prolonged speech silences tends to be positively correlated with the CER. At the same time, it may seem that replacing LSA with DCA allows us to synthesize intelligible speech in a considerably wider range with a slightly increasing CER for extremely long utterances. However, the silence rate plot shows that Tacotron2-DCA-80 cannot generate fluent speech over nearly the entire length range. Nontrivial punctuation in this case tends to manifest as long pauses in synthesized speech. In contrast, the proposed model can synthesize intelligible fluent speech over the whole range of considered utterance lengths. In this instance, the slight increase in the silence rate is caused by an increase in the number of punctuation marks in the texts. Overall, the results demonstrate that a lower degree of alignment consistency (see Table \ref{tab2}) leads to inferior performance in terms of generalization to long utterances.

\section{Conclusions}

In this paper, we have proposed an autoregressive multispeaker TTS system that can synthesize high-fidelity voice for new speakers using extremely long texts and only a few seconds of target speech without retraining the model. The proposed system consists of three key components: a neural speaker encoder, Tacotron 2 based synthesizer and universal SC-WaveRNN vocoder. We have integrated Dynamic Convolution Attention into the synthesizer and introduced a number of changes and optimizations throughout the system that have resulted in enhanced alignment consistency and ability to generalize to extralong utterances. Experiments confirm that the proposed model can synthesize intelligible fluent speech for unseen speakers using texts that are considerably longer than those used for training. Moreover, both subjective and objective evaluations highlight that the synthetic speech exhibits a high level of naturalness and similarity to the target speaker.

Overall, this research opens up the opportunity of voice cloning for long sentences and entire paragraphs using attention-based architecture, thereby extending the application boundaries of autoregressive networks. The property of generalization to long utterances can enhance the speech naturalness due to the capability of synthesizing complete text rather than separate parts. However, the proposed system still suffers from limitations caused by the inability of prosody transfer. Future work can focus on experimentation with normalizing flows to address this problem. Audio samples of synthesized and reference speech are available on the web\footnote{https://artem-gorodetskii.github.io/long-form-voice-cloning/}.

\bibliographystyle{IEEEtran}
\bibliography{mybib}

\end{document}